\documentclass[preprint]{revtex4-1}
\usepackage{graphicx}
\usepackage{dcolumn}
\usepackage{bm}
\usepackage{amsmath,amssymb,graphicx}
\usepackage{graphicx}
\usepackage{bm}
\usepackage{float,placeins}
\usepackage{float}
\usepackage{adjustbox}
\usepackage{mathptmx}
\usepackage{epstopdf,hhline,adjustbox,multirow,booktabs,longtable,makecell}
\usepackage{array}
\usepackage{booktabs,ragged2e}
\usepackage[table]{xcolor}
\usepackage{multirow}
\usepackage{hhline}
\usepackage{verbatim}
\usepackage{subfigure}
\usepackage{soul}
\usepackage{blindtext}
\usepackage{hhline}
\usepackage[symbol]{footmisc}
\usepackage[utf8]{inputenc}
\usepackage{footnote}

\usepackage[colorlinks=true,linkcolor=blue]{hyperref}
\hypersetup{
    colorlinks=true,    
    linkcolor=blue,       
    citecolor=blue,       
    filecolor=blue,        
    urlcolor=black,        
    linktoc=page          
 }
\begin{document}

\title{Gallium-Boron-Phosphide (GaBP$_{2}$): 
 A New III-V Semiconductor for Photovoltaics}

\author{Upendra Kumar}
\thanks{These two authors contributed equally}
\author{Sanjay Nayak}
\thanks{These two authors contributed equally}
\author{Soubhik Chakrabarty}
\author{Satadeep Bhattacharjee}
\author{Seung-Cheol Lee}
\email{seungcheol.lee@ikst.res.in}
\affiliation{Indo-Korea Science and Technology Center (IKST), New Airport Road, Yelahanka, Bangalore, India- 560064}
\date{\today}

\begin{abstract}
\noindent  
Using machine learning (ML) approach, we unearthed a new III-V semiconducting material having an optimal bandgap for high efficient photovoltaics with the chemical composition of Gallium-Boron-Phosphide (GaBP$_{2}$, space group: Pna2$_1$). ML predictions are further validated by state of the art ab-initio density functional theory (DFT) simulations. The stoichiometric Heyd-Scuseria-Ernzerhof (HSE) bandgap of GaBP$_{2}$ is noted to 1.65 eV, a close ideal value (1.4-1.5 eV) to reach the theoretical Queisser-Shockley limit. The calculated electron mobility is similar to that of silicon. Unlike perovskites, the newly discovered material is thermally, dynamically and mechanically stable. Above all the chemical composition of GaBP$_{2}$ are non-toxic and relatively earth-abundant, making it a new generation of PV material. Using ML, we show that with a minimal set of features the bandgap of III-III-V and II-IV-V semiconductor can be predicted up to an RMSE of less than 0.4 eV. We presented a set of scaling laws, which can be used to estimate the bandgap of new III-III-V and II-IV-V semiconductor, with three different crystal phases, within an RMSE of $\approx$ 0.5 eV. 


\end{abstract}

\maketitle
\section{ Introduction}   

Photovoltaic (PV) conversion is the direct conversion of light into electricity and thus has a true potential to replace fossil fuel based energy resources which have harmful consequences on nature. One of the major components of the PV devices is the light-absorbing material that consists predominantly
of semiconducting materials with an electronic bandgap (E$_g$) characteristic. The suitability of these PV materials is evaluated by two major properties: (i) its capability of absorption of incident light (absorption coefficient) and creation of electron-hole (e-h) pair and (ii) their ability to transport these e-h carriers from semiconducting material to electrical contacts. There are a few other criteria to follow while selecting PV material such as its toxicity, and its constituent's earth abundant etc.
\par
In their classic work, William Shockley and Hans J. Queisser
\cite{shockley1961detailed,miller2012strong} showed that for a single junction solar cell the maximum theoretical efficiency can be $\approx$ 33.5\% given that the E$_g$ of the semiconductor is $\approx$1.4-1.5 eV \cite{shockley1961detailed,gratzel2001photoelectrochemical}. Most of the commercialized PV devices that are available to us now are made of classical semiconducting materials  silicon (Si) \cite{priolo2014silicon} which has E$_g$ of 1.1 eV. The recorded efficiency of Si based solar cell is 6\% for amorphous and 25\% in single crystalline and single junction configurations \cite{green2009path}. Such a lower value of efficiency is caused by its inherent material properties of its indirect E$_g$ \cite{priolo2014silicon}. Other than Si, only a limited number of semiconductor composed of III-V and II-VI elements possess the desired value and nature of E$_g$ \cite{curtarolo2013high}, many of which contain indium (In), which has a very low earth abundance. Gallium Arsenide (GaAs), which is a binary III-V semiconductor having a direct bandgap of 1.4 eV, also have been proposed to be another prominent candidate for PV application but the maximum recorded efficiency of GaAs based solar cell \cite{greenu2015solar} is at 28.8\% which is still 4.7\% away from the theoretical limit. Further to achieve the maximum value of the theoretical limit of PV plenty of other different materials are explored in past including CdTe and perovskites. Nonetheless their efficiency is lower than the GaAs itself \cite{singh2012temperature,wesoff2016first,peters2018energy} till date.  In recent times organic-inorganic metal halide perovskite draws tremendous attention due to its high efficiency ($\approx$ 23.7\%) \cite{wang2019review,meng2018addressing}. Yet their poor material stability has hindered their commercial applications.

\par

Recent success in synthesis and stabilization of ternary and quaternary semiconductors through the cation mutation of III-V and II-IV-V semiconductors opens up a set of materials \cite{lahourcade2013structural,senabulya2016stabilization,feldberg2013growth,skachkov2016disorder,zakutayev2014experimental}. Predominantly they crystallize in three different structural phases, namely  (i) Wurtzite-kesterite (KT) (space group: Pna2$_1$) (ii) Wurtzite-stannite (ST) (space group:Pc)  and (iii) chalcopyrite (CP) (space group: I$\bar4$2d ). Recent studies show that semiconducting material composed of elements from group-II, group IV and V as well as group-III and V in ABX$_2$ form can absorb visible light \cite{wang2014design,fioretti2015combinatorial}. This further suggests that ABX$_2$ [A=\{Al, Ga, Zn, Cd\}, B=\{Ga, In, Ge, Sn, Si\} and X = \{N, P, As, Sb\}] can be a new generation materials for PV application \cite{fioretti2015combinatorial,feldberg2013growth,martinez2016solar}. One of the interesting and important properties of ABX$_2$ is their higher degrees of tunability of E$_g$ by foreign element substitution, alloying, and phase engineering \cite{wang2014design,fioretti2015combinatorial,narang2014bandgap}. 
\par
Pandey \textit{et al.}'s \cite{pandey2017ii} theoretical work suggests a few semiconducting materials with E$_g$ of $\approx$ 1.4 eV namely, ZnSnP$_2$, CdSiAs$_2$, GaInN$_2$, ZnSiP$_2$, AlInAs$_2$, CdGeP$_2$, AlInAs$_2$ etc. It is to note that toxicity is one of the biggest disadvantages for compound which contains element like arsenic and cadmium. Further, there are many reports in the literature which hinting that indium (In) has a tendency of forming metallic cluster (or segregation) inside the material. This has a detrimental effect to the device performance of In-based semiconductors\cite{tang2015indium,lee2018strain,reyes2013photoluminescence,martini2003influence,barlow2018observing,mannarino2015surface}. Further ZnSnP$_2$ which has a experimental E$_g$ of $\approx$ 1.7 eV shows a order-disorder phase transition at high temperature which reduce the E$_g$ to 0.75 eV \cite{scanlon2012bandgap}. The experimental E$_g$ of ZnSiP$_2$ in CP phase is $\approx$ 2.1 eV a bit higher for use in single junction solar cell but can be used for multijunction tandem devices \cite{martinez2016solar}. Theoretical work of Gautam \textit{et al.} shows that the higher theoretical limit of efficiency of CdGeP$_2$
based solar cell is 22.6\% \cite{gautam2015structural} which is less than the efficiency of Si based solar cell. Because of all the above issues, finding new materials in the ABX$_2$ family is essential for designing efficient PV devices.  In all of these studies mentioned earlier, the inclusion of boron in ABX$_2$ materials were neglected completely. In this regard, the property of the boron based ABX$_2$ semiconducting materials are fundamentally important too. In this work, we explore material properties such as E$_g$ of boron based III-III-V$_2$ materials and discussed its viability in PV and/or other optoelectronic applications. We substituted boron at both A and B sites of ABX$_2$ independently along with other group-III elements such as (In,Ga,Al) which gives us 24 new sets of materials in one phase alone. 
\par

Most conventional ways to find the E$_g$ of semiconducting materials are
\begin{enumerate}
	\item Optical Spectroscopy  or electrical transport characterization of chemically synthesized samples
	\item Ab-initio Density Functional Theory (DFT) simulations (Using Heyd-Scuseria-Ernzerhof (HSE) and/or GW approximation)
\end{enumerate}
Both these above mentioned techniques required a substantial amount of experimental and computational resources respectively and are time consuming process especially for large set of samples. Recent development of machine learning (ML) based approach has shown a great promise in predicting material properties \cite{butler2018machine,sparks2016data,xue2017informatics} within a fraction of a second. Gamst \textit{et al.} predicted the mechanical hardness of poly-crystalline inorganic compounds using neural network \cite{de2016statistical}. Further Zheng \textit{et al.} classify the stability of  full-Heusler compounds using convolutional neural networks (CNN) \cite{zheng2018machine}. Ramprasad \textit{et al.} predicted the various material properties such as atomization energy, crystal structure, electron affinity, dielectric constant, formation energy etc using multiple ML models \cite{pilania2013accelerating}. Motivated by the above works, we explore various supervised ML models to predict E$_g$ of ABX$_2$ compounds and used the trained model to predict bandgap of boron based compounds. We validate the ML-predicted E$_g$ of newly predicted materials using ab-initio DFT calculations where mBJ and HSE06 functional are used. We find that our ML predicted and DFT-calculated E$_g$ values are in good agreement. By imposing various criteria that is necessary for a good PV material we filtered new semiconductor for efficient PV applications.
\par
To this end, we arrange the rest of the paper in the following order;
\begin{enumerate}
	\item Section II deals with the methods of ML models and ab-initio simulations
	\item In section III we discussed the predictability of our ML models and filter new materials for PV application with suitable E$_g$
	\item We validate the E$_g$ of newly predicted material using ab-initio calculations in section IV. We further studied their structural, electronic, transport properties as well as stability and further filtered them for PV applications.
\end{enumerate}

\section{Methods}
\subsection{Dataset selection}
To train the ML models, we used the data-set generated by Pandey {\it et al.} \cite{pandey2017ii} where E$_g$ of ABX$_2$ [A=\{Al, Ga, Zn, Cd\}, B=\{Ga, In, Ge, Sn, Si\} and X = \{N, P, As, Sb\}] type stoichiometric semiconductor are computed with first-principle DFT simulations. It is well known to the literature that local and semilocal approaches severely underestimate the E$_g$ of materials. To overcome this issue, the authors included the meta-GGA GLLB-SC functional in their calculations and showed that the calculated E$_g$ is in good agreement with the experiment \cite{pandey2017ii}. In their work the author chose three different crystal structure i,e. (i) KT, (ii) ST and (iii) CP phases \cite{goodman1957new,jaffe1984theory}. Details of their computational technique and numerical parameters are discussed in Reference \onlinecite{pandey2017ii}. The data-set contains relatively a small number of entry ($\approx$ 100).
\subsection{Features selection} 
In predictions of E$_g$ of ABX$_2$ type compounds semiconductor, we include various properties of the compositional elements of ABX$_2$ as features. They are {\bf (1)} atomic number (AN), {\bf(2)} atomic mass (AM), {\bf(3)} period (P) and {\bf(4)} group in the periodic table (G), {\bf(5)} first ionization energy (FIE), {\bf(6)} second ionization energy (SIE), {\bf(7)} electron affinity (EA), {\bf(8)} Pauling electronegativity $(\chi)$, {\bf(9)} Allen electronegativity (AEN), {\bf(10)} Van der Waals radius $(r_{W})$, {\bf(11)} covalent radius  $(r_{cov})$, {\bf(12)} atomic radius  $(r_{atom})$, {\bf(13)} melting point (MP), {\bf(14)} boiling point (BP), {\bf(15)} density $(\rho)$, {\bf(16)} molar volume $(V_{m})$, {\bf(17)} heat of fusion $(\Delta_{fus}H)$, {\bf(18)} heat of vaporization $(\Delta_{vap}H)$, {\bf(19)} thermal conductivity $(\sigma_{T})$, and {\bf(20)} specific heat $\mathrm{(c_{v})}$ etc. We considered volume of the unit cell (V) also to a component to feature vector. We classify the crystal phase of the materials using one-hot encoding technique in our work. The numerical values of features are given in section I of Supplementary Information (SI).
\subsection{Machine learning methods}
\noindent In constructing the ML models, we employed various linear [e.g. Ordinary Least Square (OLS), Partial Least Square (PLS), Ridge and least absolute shrinkage and selection operator (Lasso)] and non-linear regression [e.g. Gradient Boosting Regression (GBR), Kernel Ridge Regression (KRR),  Random Forest Regression (RFR), Support Vector Machine (SVM), Artificial Neural Network (ANN)] methods via \texttt{Scikit-Learn} library \cite{scikit}. We also checked the prediction power of recently developed XG-Boost regression (XGB) technique \cite{chen2016xgboost} in this study. Details of the hyperparameters used in these models are discussed in section II of SI.  We chose 75\% of the data to train models while remaining data to test them. The Monte Carlo cross-validation method is used to evaluates their predictability. Before the machine learning regression, the feature vectors were normalized. We assessed the prediction power of ML models with help of root mean square error (RMSE) and R$^2$ value of test dataset.

\subsection{Ab-initio computational details}
Ab-initio Density Functional Theory (DFT) simulations were carried out using Vienna ab-initio simulation package (VASP)  where projector augmented wave method (PAW) was used \cite{kresse1996efficiency,kresse1996efficient,kresse1999ultrasoft}. A generalized gradient approximation proposed by  Perdew-Burke- Ernzerhof \cite{perdew1996generalized} was used for calculation of the exchange and correlation energy.  Brillouin zone is sampled on a $\Gamma$-centred (8$\times$6$\times$7 for KT, ST and 8$\times$8$\times$8 for CP phase) uniform mesh of k-points in a unit cell of reciprocal space \cite{monkhorst1976special}. A plane wave cutoff energy of 520 eV was used in our simulations. Positions of all the atoms were allowed to relax to minimize energy until forces on each atom were less than 10$^{-3}$ eV/$\AA$. To correct the known underestimation of E$_g$ by DFT-PBE, we included modified Becke-Johnson exchange (mBJ) potential and  \cite{becke2006simple} and Heyd-Scuseria-Ernzerhof (HSE) screened Coulomb hybrid functional \cite{heyd2005energy} in our calculations. In calculation with HSE functional, we used 25\% of the Hartree-Fock exchange potential. The screening parameter was fixed at 0.2. The dynamical stability of materials was studied using Density Functional Perturbation Theory (DFPT). The electronic transport properties were calculated by solving Boltzmann transport equation (BTE) as implemented in AMMCR code \cite{mandia2019ammcr}.
\par

\begin{figure}
	\centering
	\includegraphics[scale=0.50]{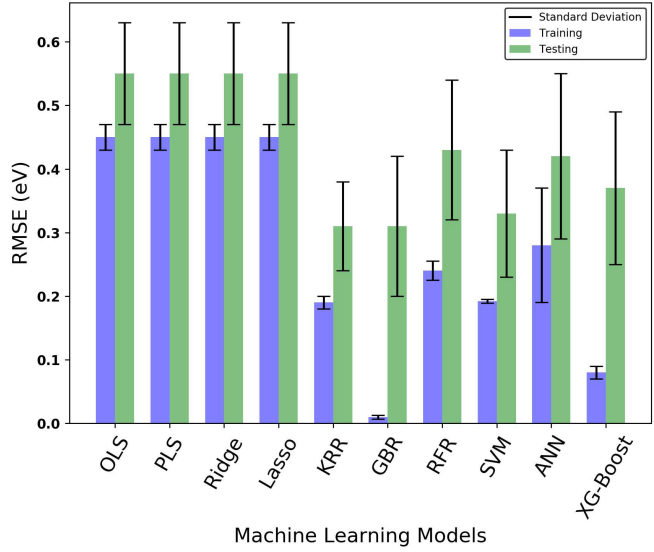}
	\caption{\small RMSE in training and testing for prediction of the bandgap, when 64 features are taken into consideration.}
	\label{feature64}
\end{figure}

\begin{figure*}
	\centering
	\includegraphics[scale=1.0]{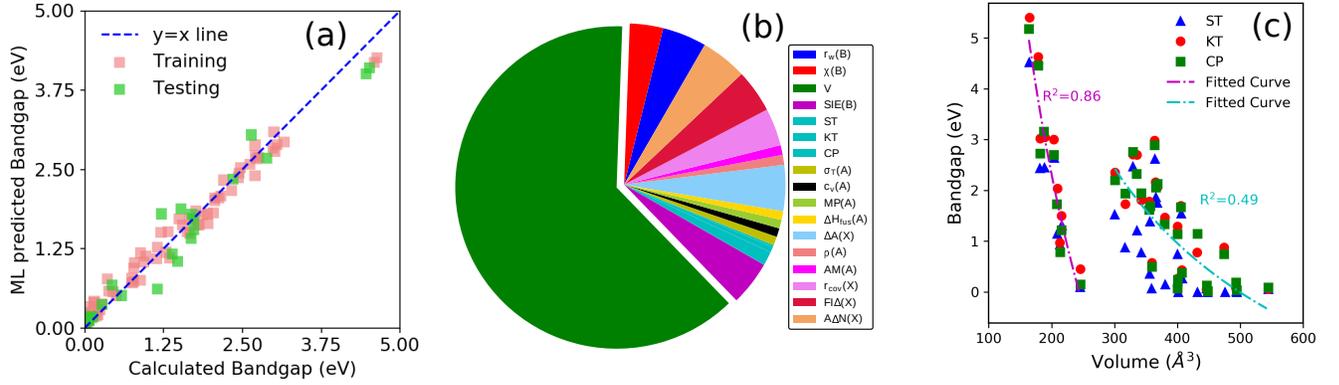}	
	\caption{\small (a) shows the comparison between training data and testing data with Random Forest Regression in the prediction of E$_g$.(b) shows the weightage of top 17 features in the prediction of GLLB-SC E$_g$ of ABX$_2$ compounds. (c)shows the relationship between E$_g$ and the volume of the unit cell.}
	\label{weightage}
\end{figure*}
\vspace{0.1in}
\section{Machine Learning Formulation and Screening of Photo-Voltaic Materials}
We begin our prediction of E$_g$ using the features as discussed in section II. The predictability of different models are compared (see Fig.\ref{feature64}) and we find that predictability of non-linear models are better comparison to linear models (i.e. low RMSE in predictions and better R$^2$) in prediction of $E_{g}$. The RMSEs (standard deviation (SD)) in training and testing cases in various linear models are essentially the same and are 0.45 (0.02) and 0.55 (0.08) eV respectively. Non-linear model like KRR, GBR, RFR, SVM, ANN and XGB predicts the E$_g$ up to an RMSE(SD) of 0.31(0.07), 0.31(0.11), 0.43 (0.11), 0.33 (0.12), 0.42 (0.13) and 0.37 (0.12)  eV respectively. We further compared our prediction accuracy with the recently published work in the literature.  Recent work of Huang \textit{et al.} predicts the E$_g$ of binary nitrides material with SVR up to an RMSE of 0.298 eV \cite{huang2019band}. By using a relatively larger dataset of MXenes materials Ranjan \textit{et al.} showed a better prediction accuracy (test RMSE of $\approx$ 0.20 eV) \cite{rajan2018machine}. Olsthoorn \textit{et al.} predicted the E$_g$ of organic crystal up to an RMSE of $\approx$ 0.5 eV \cite{olsthoorn2018band}. The E$_g$ predicted by ML for double perovskite materials are within the RMSEs of 0.8-1.0 eV \cite{pilania2013accelerating}. Similarly, the work of Sotskov \textit{et al.} \cite{sotskov2018band} showed a relatively low accuracy in predictions of E$_g$ of inorganic materials. Thus, in general our prediction accuracy of E$_g$ in ABX$_2$ materials are within the error broadly reported in the literature which further suggests, we have achieved to build a working ML model with the optimal set of hyperparameters.

\par
Based on the important features in the prediction of the E$_g$ we minimized the number of feature to 17, which includes one-hot encoding of the phase of materials. Feature weightage are listed in section III of SI. The relative weightage of all 17 features are shown in a pie-chart plot as Fig.\ref{weightage}(b) when predictions are made with top 17 features only. We find very little change in the prediction error in comparison to the 64 features. We note that the volume of the unit cell (V) is the predominant factor that determines the E$_g$ of ABX$_2$ materials. (see Fig.\ref{weightage}(b)). Fitting of the empirical relation between E$_g$ and square of lattice parameter (or V$^{2/3}$ in our model) \cite{PhysRevB.8.6033} to the data reveals two different classes and hence two sets of coefficient (see Eqn.\ref{egvsv}). We find that when the volume is less than 246 $\AA^3$ (i.e. for ABN$_2$) the E$_g$ increases quite rapidly with decrease in the V. Compound having anions as P, As, and Sb shows a rather slower changes in the E$_g$ with changes in V. Statistically, it can be observed that a non-nitride ABX$_2$ material with unit cell volume of 300-420 $\AA^3$ can have a band-gap of $\approx$ 1.4-1.5 eV which is ideal for PV applications.  

\begin{widetext}
	
\begin{equation}
\centering
E_{g}= 
\begin{cases} 
-16.54 +\frac{622.51}{V^{2/3}} & V \leq 246 \ \AA^3\quad  \mathrm{ABN_{2}\;\; family \;\;  : \;\;R^2 =0.86}\; \mathrm{\; and\; RMSE\; = 0.54 \; eV}\\

 
-6.10 +\frac{368.17}{V^{2/3}} & V \geq 290 \ \AA^3 \quad  \mathrm{Others\;\; : \;\; R^2 =0.49}\; \mathrm{\; and\; RMSE\; = 0.47 \; eV}

\end{cases}  
\label{egvsv}
\end{equation} 
\end{widetext}

\begin{figure} [hb]
	\centering

		\includegraphics[scale=0.70]{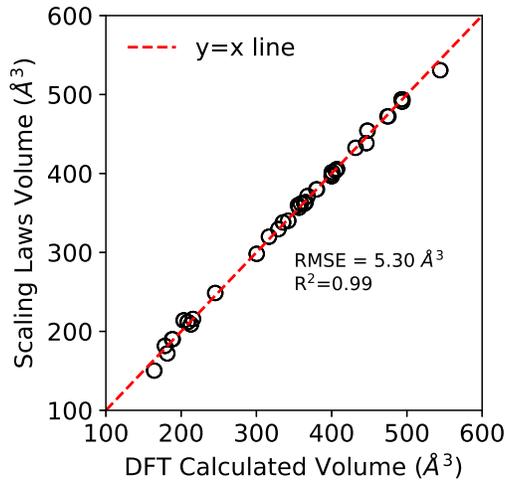}

	\caption{\small Comparison between volume calculated with Eqn. \ref{volume_vs_feat} and DFT computed of the unit cell.}
	\label{VolumeTest}
\end{figure} 

 \par
It is quite clear by this point that prediction of E$_g$ with ML, V is not only important but also necessary. As volume can only be calculated by only DFT or experimentally (X-ray diffraction method), without a further ML model for V, our previous model for E$_g$ will be not effective. Thus, we designed a ML model to predict the V. Interestingly, with the LASSO regression, we predict V quite accurately (up to an RMSE of 5.3 $\AA^3$) (see Fig.\ref{VolumeTest}). Based on which we establish a relation between V and atomic features (see Eqn.\ref{volume_vs_feat}) of the constituent elements;
 
{\centering
\begin{align}
V=\Big[-573.10+1.76(r_{cov})_{A}+2.00(r_{cov})_{B} +4.08(r_{cov})_{X}\Big]\;\AA^{3}
\label{volume_vs_feat}
\end{align}} 
\noindent where $r_{cov}$s are in picometer(pm) unit.

\par
\begin{table*}
	
	\caption{ \small machine learning (ML) predicted material properties with various regression techniques vs. DFT computed values. The $\pm$ sign is for standard deviation in our predictions. All mentioned values are for wurtzite-kesterite (KT) Phase.}
	\label{tab-resultsxxx}
	\centering
	\resizebox{0.95\textwidth}{!}{
		\begin{tabular}{|c|c|c|c|c|c|c|c|c|c| lllllllll@{}} 
		\hline				
		Compound  & Scaling- & DFT-PBE  & E$_g$  & ML-E$_g$ & ML-E$_g$ & DFT-mBJ  & DFT-HSE06  & DFT-PBE  \\
		Name & Volume ($\AA^3$) &  Volume ($\AA^3$) &  (Eqn.\ref{egvsv}) (eV) & (RFR) (eV) & (XGB) (eV) &  E$_g$ (eV) &  E$_g$ (eV) & $\Delta H_f $ (eV/atom)\\ \hline

		AlBAs$_2$     & 290.0    & 293.5          & 2.30            & 1.47 $ \pm 0.14$            & 1.53 $\pm 0.15$            & 2.63            & -                  & -                           \\ \hline
		AlBP$_2$      & 240.8    & 253.7          & 3.41            & 2.09  $\pm 0.14$            & 1.61 $\pm 0.20$            & 1.65            &     -              &   -0.35                         \\ \hline
		BAlP$_2$      & 250.1     & 253.6          & 3.17            & 2.4 $\pm 0.09$            & 2.63 $\pm 0.14$            & 1.65            & 1.83                 &   -0.35                        \\ \hline
		BAlSb$_2$     & 380.7    & 380.3          & 0.91            & 1.59 $\pm 0.13$           & 2.03$\pm 0.18$            & 0.63            &     -              &   -                         \\ \hline
		BGaSb$_2$     & 382.7    & 376.4          & 0.88           & 1.31 $\pm 0.14$           & 1.5$\pm 0.15$            & -            &   0.92                &   -                         \\ \hline
		GaBP$_2$      & 242.4   & 253.5          & 3.37             & 1.51 $\pm 0.08$            & 1.56$\pm 0.16$            & 1.50            &   1.65                &    -0.27                        \\ \hline
		InBP$_2$      & 277.8    & 286.6          & 2.54            & 1.47  $\pm 0.08$          & 1.51$\pm 0.20$            & 1.20            &      1.55             &     -0.09                       \\ \hline
		
	\end{tabular}}
	
\end{table*}

\par
Further, we extend our work to predict the E$_g$ of ABX$_2$ for boron at A and B sites independently by which we generate 24 new configurations. At first, we predict their volumes using Eqn.\ref{volume_vs_feat}. Based on our initial assessment, we find that inclusion of boron in ABN$_2$ material predicts a low V and hence a high E$_g$. Thus we focused only on its non-nitride forms. We find there are $7$ compound which posses the desired range of volume which can have E$_g$ of $\approx$ 1.4-1.5 eV (see Table \ref{tab-resultsxxx}). We predicted E$_g$ of these new materials using Eqn.\ref{egvsv} and various ML models that we constructed (see Table \ref{tab-resultsxxx}). There is a quite mismatch in their prediction which we will address in coming section. We find using ML that GaBP$_2$ full filled our desired criteria {\it i.e.} it does not contain any toxic element(s) and indium free but most importantly E$_g$ are close to 1.4-1.5 eV. At this stage we ignore BAlSb$_{2}$ and BGaSb$_{2}$ from the list as antimonides show positive formation enthalpy \cite{pandey2017ii}. In our analysis, we included BAlP$_2$ and AlBP$_2$ as XGB shows one of them might have bandgap near to 1.6 eV.

\section{Ab-initio Validation}

The ML-volumes estimated from Eqn.\ref{volume_vs_feat} are compared with the DFT-PBE volumes (see Table \ref{tab-resultsxxx}). We find that ML predicted volumes are in good agreement with the DFT-PBE computed ones. We assess the E$_g$ of the probable compound using DFT-mBJ functional as it is relatively less computationally expensive.  We compared the E$_g$ estimated with Eqn.\ref{egvsv} and DFT-mBJ. We find there is a large mismatch between them especially when the V in range of 240 to  290 $\AA^3$. Thus, we propose a new scaling relation (see Eqn.\ref{egvsv1}) between E$_g$ and V in this region.  
 
\begin{align}
E_{g}&=& 
-3.50 +\frac{204.2}{V^{2/3}}\;\; \mathrm{for}\;\; 240\;\;<\; \mathrm{V}\;\;< 290 \ \AA^3 \nonumber \quad
\\&&{R^2 =0.80}\; \mathrm{\; and\;\; RMSE\; = 0.07\; eV}
\label{egvsv1}
\end{align} 
Now together Eqn.\ref{egvsv} and Eqn.\ref{egvsv1} represent a full set of scaling relations between E$_g$ and V for ABX$_2$ materials. 
\par
Further, we find that AlBP$_2$ and BAlP$_2$ have DFT-mBJ E$_g$ of 1.65 eV which suggests our ML values in this case are is bit overestimated. Nevertheless, our finding of the ML and DFT-mBJ E$_g$ of GaBP$_2$ is in good agreement. To obtain the exact E$_g$ we used DFT-HSE06 functional which revealed AlBP$_2$ has E$_g$ of 1.83 eV (see Table \ref{tab-resultsxxx}). This value is relatively higher for single junction PV applications, hence we exclude it for further analysis. The obtained HSE06  E$_g$ of the GaBP$_2$ is 1.65 eV, a close ideal case for high efficiency PV material. Thus, we discussed the material properties of GaBP$_2$ in details. The material stability and viability of its use in PV are also discussed thoroughly in the rest of the paper.

\subsection{Atomic and Electronic Structure}
\begin{figure}[htb]
	\centering
	\includegraphics[scale=0.55]{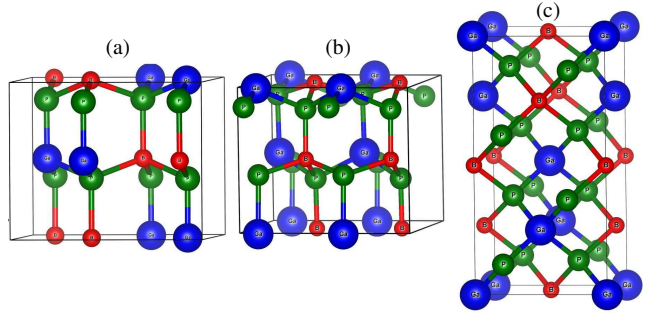}
	
	\caption{\small Ball and stick model of GaBP$_2$ in  Wurtzite-kesterite (KT) (\textbf{a})  Wurtzite-stannite (ST) (\textbf{b}) and chalcopyrite (CP) (\textbf{c})
		crystal phase. Blue, green and red ball represent gallium (Ga), phosphorus (P) and boron (B) atoms respectively.  }
	\label{structure}
\end{figure}
We obtain relaxed atomic structure and total energy of GaBP$_2$ in three different phases (i.e. KT, ST and CP (see Fig.\ref{structure})). Comparison of their formation enthalpy ($\mathrm{\Delta H_f}$) using Eqn. \ref{enthalpy} suggests KT and CP phases are more energetically favourable than ST phase.
\begin{equation}\label{enthalpy}
\mathrm{\Delta H_f (ABX_2)= E_{tot} (ABX_2)-\mu(A)-\mu(B)-2\times \mu(X)}
\end{equation}
where $\mathrm{E_{tot} (ABX_2)}$ is the total energy of formula unit cell of $\mathrm{ABX_2}$ and $\mu(i)$ represent the chemical potential of $i$ element (For details see section IV of SI). The negative value of $\mathrm{\Delta H_f}$ (see Table \ref{tab-resultsxxx}) suggests that the material can be synthesized at thermal equilibrium conditions.

\subsubsection{KT Phase}
\begin{figure}
	\centering
	\includegraphics[scale=0.5]{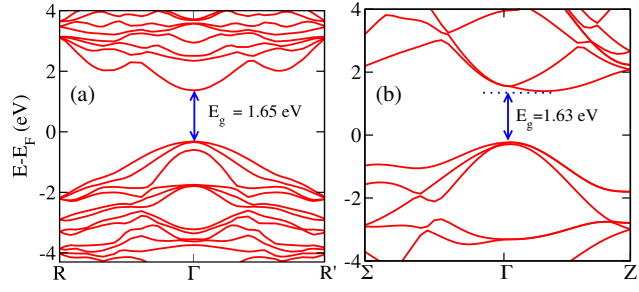}
	
	\caption{\small HSE06 bandstructure of GaBP$_2$ in Wurtzite-kesterite (KT) (a) and chalcopyrite (CP) (b) phases.}
	\label{band}
\end{figure} 

\par

The DFT-PBE optimized lattice constant for GaBP$_2$ are $a$ = 6.07 \AA,  b = 7.13 \AA \space and c = 5.85 \AA  \space and $\alpha$ = $\beta$ = $\gamma$ = 90$^\circ$ is noted. The unit cell contains 16 atoms. At the basal plane Ga-P bond-length is 2.36-2.39 \AA \space while axial bond is 2.38 \AA \space long. Similarly, the planar and axial bond of B-P is 2.00 \AA \space and 2.02 \AA \space long respectively. We note the unit cell volume to be 253.5 \AA$^3$. Electronic structure obtained with DFT-PBE revealed a direct E$_g$ of 0.97 eV at $\Gamma$ point (see section V of SI for band and dos plot). Atom and orbital projected densities of states (DOS) further revealed that the valence band (VB) is dominated by $p$-orbital of P, a small contribution form $p$-orbitals of Ga and B with a minute contribution from $s$-orbital of P and Ga. $s$-orbital of B appears $\approx$ at -5.0 eV below valence band maximum (VBM). The conduction band (CB) of the electronic structure is majorly contributed by Ga-$s$, Ga-$p$, B-$p$, P-$p$, and P-$s$ orbitals. A small contribution from Ga-$d$ and B-$s$ orbital is also noted ( see Fig. S2 (a) of SI).
\par
It is well known to the literature that DFT-PBE severely underestimate E$_g$ of materials. To overcome this issue various scheme has been proposed in past \cite{morales2017empirical}. In, recent times it was suggested that use of mBJ correction to the local/semi-local functional (such as LDA or GGA) can results in an estimation of accurate electronic E$_g$ \cite{chan2010efficient}. Nevertheless, use of hybrid functional (such as HSE06) along with DFT-PBE is now considered to be a gold standard for ab-initio estimation of E$_g$ \cite{garza2016predicting,nguimdo2015density}. Here we used both functional to calculate E$_g$ of materials. We found that the calculated E$_g$ of GaBP$_2$ in KT phase using mBJ and HSE06 functional is 1.50 and 1.65 respectively (see Table \ref{tab-resultsxxx}) close to our ML prediction. We present the HSE06 band-dispersion of GaBP$_2$ around $\Gamma$ point in Fig.\ref{band} (a) where a direct E$_g$ at $\Gamma$ point is noted. A full DFT-PBE band-dispersion relation is presented in Fig. S1(a) in SI. This direct E$_g$ would allow for efficient absorption of the incident photon and hence a better PV efficiency.

\subsubsection{CP Phase}

The optimized unit cell lattice constant of GaBP$_2$ in CP phase are $a$=b=c=6.07 \AA \space and $\alpha$ =  $\beta$ = 130.44$^\circ$ and \mbox{$\gamma$ = 72.70 $^\circ$}. Here, unit cell contains 8 atoms and volume of the unit cell is noted as 126.56 $\AA^3$. While the Ga-P bond is 2.37 \AA \space long, the B-P bond-length is 2.01 \AA.  Electronic structure computed with DFT-PBE shows an in-direct E$_g$ of 0.95 eV along $\Gamma$-Z point. While the valence band maximum (VBM) is at $\Gamma$  [(0,0,0)] point, conduction band minimum (CBM) is at ($0.13.\bf{b_1}$, $0.13.\bf{b_2}$, $-0.13.\bf{b_3}$) \textbf{k}-point, where $\bf{b_i}$ are the reciprocal lattice vector.  DOS analysis suggest that the VB is dominated by p-orbital of P with a contribution from $p$-orbitals of B and Ga. The CB is dominated by the $p$-orbital of P and $s$, $p$, and $d$-orbital of Ga and $p$-orbital of B (see Fig. S2(b) in SI). We find a bit smaller E$_g$ in CP phase in comparison to KT phase using HSE06 functional. We note an indirect E$_g$ of 1.63 eV ( see  Fig. \ref{band} (b)).
 
Thus, our electronic structure analysis suggests that GaBP$_{2}$ in KT-phase is most suitable for PV applications.
\subsection{Dynamical stability, Free energy and Mechanical stability}
\begin{figure*}
	\centering
	\includegraphics[scale=0.8]{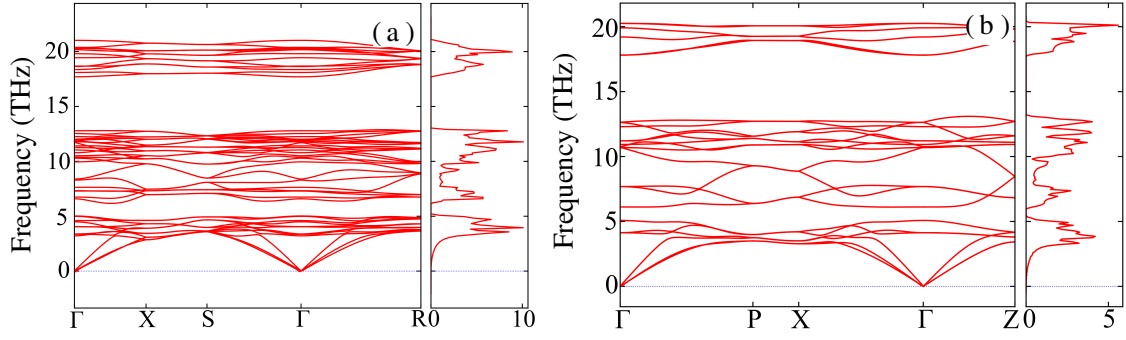}	
	\caption{\small Shows phonon dispersion of GaBP$_2$ in KT (a) and CP (b) phases respectively.}
	\label{phonon}
\end{figure*}

\begin{figure}
	\centering
	\includegraphics[scale=0.69]{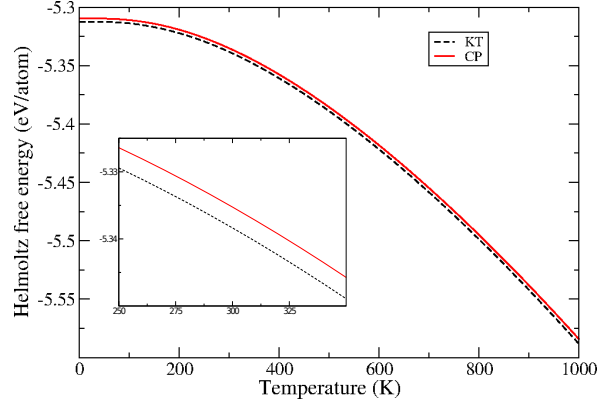}	
	\caption{ Helmholtz free energy (A) vs. temperature (T) is presented in Figure (c). Inset of (c) is the enlarged view of A vs T plot in the region of 250 to 350K.  }
	\label{stability}
\end{figure}

Further, we studied the dynamical stability of both KT and CP phase using DFPT simulation. We used phonopy code \cite{togo2015first} where a $2 \times 2 \times 2$ supercell was used for simulation of phonon dispersion. Absence of any imaginary modes (see Fig.\ref{phonon} (a) and (b)) in both structure confirmed that GaBP$_2$ can be synthesized in both phases. We estimated Helmholtz Free energy (A) as a function of temperature (T) of both configurations using the following relation:
\begin{equation}
A=E_{DFT}+F_{vib.}(T)
\end{equation} 
Where $E_{DFT}$ and $F_{vib.} (T)$ are the DFT free energy and vibrational free energy respectively. The $F_{vib.} (T)$ is calculated as \cite{togo2015first}:
\begin{equation}
F_{vib.}(T) = \dfrac{1}{2}   \sum_{qj}{\hbar\omega_{qj}}+ k_BT\sum_{qj}{}ln[1-exp(-\hbar\omega_{qj}/k_BT)]
\end{equation}
where $T$ and $K_B$ are the absolute temperature and the Boltzmann constant, respectively. $\hbar$ is the Planck's constant and $\omega_{qj}$ is the frequency of the vibration of $(q, j)$ phonon mode. $q$ and $j$ are phonon wave vector and index of the band respectively. Our estimated $A$ suggests that in all temperature region, the KT phase is more stable than the CP phase, but with a relatively small difference of around 3 meV at 300K. 
\par
We evaluate the mechanical stability by using the Born criteria \cite{mouhat2014necessary}. For KT phase which belongs to the orthorhombic crystal system, the necessary criteria are given as;
\begin{enumerate}
	\item $C_{11} \ > \ 0$
	\item $C_{11}\times C_{22} \ > C_{12}^2$
	\item $C_{11}\times C_{22}\times C_{33} \ + \ 2C_{12}\times C_{13}\times C_{23} \ - \ C_{11}\times C_{23}^2 \ - \ C_{22}\times C_{13}^2 \ -  \ C_{33}\times C_{12}^2 \ > 0$
	\item $C_{44} \ > 0$
	\item $C_{55} \ > 0 $
	\item $C_{66} \ > 0 $	
\end{enumerate}
while for the CP phase which belongs to the tetragonal crystal system, the criteria are; 
\begin{enumerate}
\item $C_{11} - C_{12} \ > 0$
\item  $2\times C_{13}^2 \ < C_{33} (C_{11} + C_{12})$
\item $C_{44} \ > 0$
\item $C_{66} \ > 0$
\item $2\times C_{16}^2 \ < C_{66} \times (C_{11}-C_{12})$.
\end{enumerate}

Coefficient obtained with DFT-PBE satisfied all of the above mentioned criteria, which suggest GaBP$_2$ is mechanically stable in both KT and CP phases. We have listed all the coefficient of stiff tensor in section VI of SI. We list other parameters corresponding to mechanical properties along with structural and electronic ones, of both phases in Table \ref{properties}.

\begin{table} 
	\caption{\small Calculated material properties of GaBP$_2$ in both KT and CP phases.}
	\label{properties}
	\begin{tabular}{l|l|l}
		\hline
		\hline
		Parameters                        & KT-Phase              & CP-Phase              \\ \hline
		$\varepsilon_s$                   & 11.95$\varepsilon_0$  & 12.10$\varepsilon_0$  \\
		$\varepsilon_\infty$              & 10.76$\varepsilon_0$  & 10.84$\varepsilon_0$  \\
		E$_D$ (eV)                        & 8.55                  & 10                    \\
		E$_g$  (eV)                           & 0.97(PBE), 1.65 (HSE) & 0.95(PBE), 1.63 (HSE) \\
		$\omega_{PO}$ (THz)               & 21.02                 & 20.25                 \\
		$C_l$ $(10^{10}N/m^2)$            & 22.43                 & 22.51                 \\
		$C_t$ $(10^{10}N/m^2)$            & 8.38                  & 8.42                  \\
		P                                 & 0.02                  & 0.02                  \\
		$\rho \ (g/cm^3)$ & 3.73                  & 3.74                  \\
		K $(10^{10}N/m^2)$                & 11.255                & 11.287                \\
		G $(10^{10}N/m^2)$                & 8.378                 & 8.419                 \\
		Y $(10^{10}N/m^2)$                & 20.137                & 20.226                \\
		$\sigma$                          & 0.202                 & 0.201                \\ \hline 
	\end{tabular}
	\par 
	\vspace{1ex}
	{\raggedright $\varepsilon_s$ = low frequency dielectric constant, $\varepsilon_\infty$ = high frequency dielectric constant, E$_D$ = acoustic deformation potential, $\omega_{PO}$ = Polar optical phonon frequency for the longitudinal mode, $C_l$ = longitudinal elastic constant, $C_t$ = transverse elastic constant, P = dimensionless piezoelectric coefficient, $\rho$ = density, K = Bulk modulus, G = Shear modulus, Y = Young modulus, 	$\sigma$ = poisson's ratio \\  \par }
\end{table}

\subsection{Electronic Transport}
 In addition to electronic structure, we calculated the mobility of n-type GaBP$_2$ in both KT and CP phase using AMMCR code \cite{mandia2019ammcr}. In our transport calculations, we included four scattering mechanisms viz. (i) ionized impurity, (ii) polar optical phonon (POP), (iii) acoustic deformation potential and (iv) piezoelectric scattering. It is worth mentioning here that at 300K, POP scattering has a significant influence on the transport properties of the III-V semiconductors. The fact that POP scattering is inelastic and anisotropic, makes relaxation time approximation inappropriate \cite{rode1975low} for calculation the semi-classical transport properties of the III-V semiconductors. In this work, we solved the Boltzmann transport equation (BTE) using Rode’s iterative method \cite{rode1975low} in order to obtain the perturbation in the electron distribution function due to different scattering mechanisms. Input parameters required for transport calculation viz. bandgap, optical phonon frequency, low and high frequency dielectric constants, dimensionless piezoelectric coefficient, the group velocity of the electron in the conduction band, elastic constants etc. are computed using DFT-PBE/DFPT-PBE and are tabulated in Table \ref{properties}. Details of the methodology of our transport calculations are discussed in references \onlinecite{mandia2019ab,faghaninia2015ab}. In KT phase the electron mobility is relatively higher compared to the CP phase thereby making the KT-GaBP$_2$ more suitable for PVs and other optoelectronic applications. At 300K the estimated mobility of the KT and CP phases with a donor concentration of 10$^{17}$ cm$^{-3}$ is $\approx$ 1340 and $\approx$ 960 $\mathrm{cm^2/V.s}$, respectively. For temperature (T) dependence of mobility in KT phase refer to section VII of SI. We note that in the T range of 150-700K, mobility decreases with increase in T while in low temperature region (50-150K) it increases with the increase in T (see Fig. S3 of SI). 
\par

Thus, our analysis from the ab-initio simulations suggests that GaBP$_2$ is most stable in KT phase with a direct E$_g$ of 1.65 eV, an excellent condition for the fabrication of high efficient PV solar cell. Negative formation enthalpy, dynamical and mechanical stability analysis together suggest that GaBP$_2$ can be chemically synthesizable and is stable in KT phase. We find that energetic and properties of CP phase are very similar to the KT phase except the nature of the electronic E$_g$. 
\vspace{-0.8cm}   
\section{Conclusion}
In conclusion, we designed machine learning models to predict the bandgap of ternary II-IV-V and III-V semiconductors in ABX$_2$ phase using a small and freely accessible dataset. The estimated RMSEs in predicting E$_g$ is less than 0.4 eV. We obtained a set of numerical scaling laws for estimation of E$_g$ using the unit cell volume as a single feature, which further concludes the compositional clustering of data between nitrides and non-nitrides. We filtered a new III-III-V$_2$ semiconducting material with the chemical composition of GaBP$_2$ which is suitable for PV applications. We predicted the electronic bandgap of GaBP$_2$ and others using ML and validate them using ab-initio numerical simulations with mBJ and HSE06 functional. We studied the structural and electronic properties, and their thermal, dynamical and mechanical stability and concluded that the newly discovered material is stable. The estimated electron mobility of the GaBP$_2$ is very similar to that of Si. The elemental earth abundance of the constituent elements, the electronic bandgap and electron mobility value along with a stable structural phase clearly suggest GaBP$_2$ will be a next generation material for photovoltaic applications.  
\section*{Data Availability Statement}
\noindent The data that support the finding of this study are available from corresponding author upon reasonable request.    
\section*{author contribution}
\noindent UK and SN conceived the idea and contribute equally in the the machine learning and DFT 
part of the calculation. SC carried out the transport calculations. SN and UK wrote the mansucript and read by all authors. SB and S-C. L supervised the project. 
\section*{Conflicts of Interest}
There are no conflicts to declare.
\bibliographystyle{apsrev4-1}
\bibliography{MachineLearning}

\end{document}